\documentclass{PoS}

\newcommand{\pslash}{p\kern-1ex /}
\newcommand{\lslash}{l\kern-1ex /}
\newcommand{\sslash}{s\kern-1ex /}
\newcommand{\Dslash}{{\cal D}\kern-1.5ex /}

\newcommand{\beqa}{\begin{eqnarray}}
\newcommand{\eeqa}{\end{eqnarray}}
\title{$\theta$ vacuum physics from QCD at fixed topology}

\ShortTitle{$\theta$ vacuum physics from QCD at fixed topology}

\author{Sinya Aoki\\
  Graduate School of Pure and Applied Sciences, 
  University of Tsukuba, Tsukuba 305-8571, Ibaraki Japan\\
  Riken BNL Research Center, 
  Brookhaven National Laboratory, Upton, NY 11973, USA\\
        E-mail: \email{saoki@het.ph.tsukuba.ac.jp}}

\author{Hidenori Fukaya\\
  Theoretical Physics Laboratory, RIKEN, Wako 351-0198, Japan\\
        E-mail: \email{hfukaya@riken.jp}}

\author{Shoji Hashimoto\\
  High Energy Accelerator Research Organization (KEK), 
  Tsukuba 305-0801, Japan\\
  School of High Energy Accelerator Science,
  The Graduate University for Advanced Studies (Sokendai),
  Tsukuba 305-0801, Japan\\
        E-mail: \email{shoji.hashimoto@kek.jp}}

\author{\speaker{Tetsuya Onogi}\\
  Yukawa Institute for Theoretical Physics, 
  Kyoto University, Kyoto 606-8502, Japan\\
        E-mail: \email{onogi@yukawa.kyoto-u.ac.jp}}

\abstract{
We propose a method to obtain physical quantities in the $\theta$ vacuum
 from those at fixed topology, which are different by finite size
 effects. Extending the work by Brower {\it et al.}, we derive the formula 
 to estimate these finite size corrections for arbitrary correlators 
 in terms of the topological susceptibility and the $\theta$ dependence.
 Applying this formula, we show that topological
 susceptibility can be measured through two-point functions of
 pseudoscalar operator. }

\FullConference{The XXV International Symposium on Lattice Field Theory\\
		 July 30-4 August 2007\\
		 Regensburg, Germany}

\begin{document}

%-------------------------------------------
\section{Introduction}
%-------------------------------------------

In Quantum Chromodynamics (QCD), the ground state must be the $\theta$
vacuum in order to ensure the cluster decomposition
property of physical observables. However, in unquenched lattice QCD
simulations by Hybrid Monte Carlo algorithm~\cite{Duane:1987de},
the correct sampling of topological charge will become increasingly more
difficult~\cite{Alles:1996vn, Del Debbio:2004xh}.
In view of this situation, one promising approach is to fix the
topology during the Hybrid Monte Carlo simulation and try to extract
physics from the simulations at a fixed topological charge $Q$.
It is known that the fixed Q effect is a finite size effect which
vanishes in the infinite volume limit. In this report, we give a
theoretical basis for estimating the finite size effect in order to extract
physics in the $\theta$ vacuum from QCD at fixed
topology~\cite{Aoki:2007ka} , by extending the work by Brower et
al.~\cite{Brower:2003yx}.  
We derive a general formula to estimate this relation.
Furthermore, using our formula, we propose a method to measure the
topological susceptibility at fixed topology. The numerical study is
given in a separate reports~\cite{Aoki:2007pw, Chiu}.

% ------------------------ section 2 ----------------------
\section{General formula}
\label{sec:general_formula}
% ------------------------
%
Consider the partition function in the $\theta$ vacuum defined by
\begin{eqnarray}
  Z(\theta) &\equiv& \langle \theta \vert \theta \rangle 
  = \exp[-VE(\theta)], 
\end{eqnarray}
where $E(\theta)$ is the vacuum energy density.
The topological susceptibility $\chi_t$ at $\theta=0$ is defined by
\begin{equation}
  \chi_t = \frac{\langle 0\vert Q^2\vert 0\rangle}{V}
=\left.\frac{d^2 E(\theta)}{d\theta^2} \right\vert_{\theta=0}.
\end{equation}
Since $\chi_t \ge 0$ by definition, $\theta =0$ is a local minimum of
$E(\theta)$. 
Moreover, Vafa and Witten proved that $Z(0) > Z(\theta)$ ~\cite{Vafa:1984xg}, 
so that $\theta =0$ is the global minimum of the function $E(\theta)$.
Assuming analyticity of  $E(\theta)$ near $\theta=0$, 
we can expand $E(\theta)$ as
\begin{equation}
  E(\theta) = \sum_{n=1}^\infty  \frac{c_{2n}}{(2n)!} \theta^{2n} 
  = \frac{\chi_t}{2}\theta^2 + O(\theta^4).
\end{equation}

The partition function at a fixed topological charge $Q$ is a Fourier
transformation of $Z(\theta)$
\begin{equation}
  \label{eq:Z_Q}
  Z_Q = \frac{1}{2\pi}\int_{-\pi}^{\pi}\!d\theta\, Z(\theta) \exp(i\theta Q) 
  = \frac{1}{2\pi}\int_{-\pi}^{\pi}\!d\theta\,\exp( -V F(\theta)),
\end{equation}
where $F(\theta) \equiv E(\theta)-i\theta Q/V$.
For a large enough volume, we can evaluate the $\theta$ integral in
(\ref{eq:Z_Q}) by the saddle point expansion.
The saddle point $\theta_c$ is given by
\begin{equation}
  \theta_c = i \frac{Q}{\chi_t V}(1+O(\delta^2)) ,
\end{equation}
where $\delta\equiv Q/(\chi_t V)$.
We then expand $F(\theta)$ as
\begin{equation}
  F(\theta) = F(\theta_c) + \frac{E^{(2)}}{2} (\theta_c) (\theta - \theta_c)^2
  +\sum_{n=3}^\infty \frac{E^{(n)}(\theta_c)}{n!} (\theta - \theta_c)^n,
\end{equation}
where $E^{(n)}$ is the $n$-th derivative of $E(\theta)$ with respect to
$\theta$ at $\theta=\theta_c$, and is given by 
\begin{eqnarray}
  V F(\theta_c) = \frac{Q^2}{2\chi_t V}( 1+ O(\delta^2)), 
&&  E^{(2)}(\theta_c) = \chi_t  ( 1+ O(\delta^2)), \\
  E^{(2n)}(\theta_c) = c_{2n} ( 1+ O(\delta^2)), 
&&  E^{(2n-1)}(\theta_c) = \theta_c  c_{2n}(1+ O(\delta^2)).
\end{eqnarray}
By a change of variable $s = \sqrt{ E^{(2)} V }(\theta - \theta_c)$
we can rewrite the integral as
\begin{eqnarray}
  Z_Q &=&
  \frac{e^{-VF(\theta_c)}}{2\pi\sqrt{E^{(2)}V}}
  \int_{-\sqrt{E^{(2)}V}(-\pi-\theta_c)}^{\sqrt{E^{(2)}V}(\pi-\theta_c)}
  ds\,  
  \exp\left[ -\frac{s^2}{2} - \sum_{n=3}\frac{E^{(n)}V}{n!}
    \left(\frac{s}{\sqrt{E^{(2)}V}}\right)^n\right] 
\end{eqnarray}
Neglecting exponentially suppressed terms and expanding in powers of $1/V$,
we obtain 
\begin{eqnarray}
  Z_Q &=& 
  \frac{1}{\sqrt{2\pi\chi_tV}} \exp\left[-\frac{Q^2}{2\chi_t V}\right]
  \left[1 -\frac{c_4}{8 V \chi_t^2} + O\left(\frac{1}{V^{2}}, \delta^2 \right)
  \right].
\end{eqnarray}
This shows that, as long as $\delta\ll 1$ (equivalently, $Q \ll \chi_t V$), the
distribution of $Q$ becomes the Gaussian distribution.
Similarly, consider an arbitrary correlation function in $\theta$
vacuum and at fixed topological charge $Q$ are defined as
\begin{eqnarray}
G(\theta)=\langle \theta \vert  O_1 O_2 \cdots O_n \vert \theta\rangle, 
&&
G_Q = \frac{1}{Z_Q}\frac{1}{2\pi}\int d\,\theta Z(\theta)\,
 G(\theta)\exp(i\theta Q).
\end{eqnarray}
Using the saddle point expansion as before, 
if $G$ is CP-even, we can show 
\begin{eqnarray}
  G_Q^{\mathrm{even}}
  &=& G(0) +G^{(2)}(0)\frac{1}{2\chi_t V}\left[1-\frac{Q^2}{\chi_tV}-\frac{c_4}{2\chi_t^2V}\right]
  + G^{(4)}(0)\frac{1}{8\chi_t^2 V^2} + O(V^{-3}), \nonumber \\
  \label{eq:even}
\end{eqnarray}
where $G^{(n)}(0)$ stands for the n-th derivative of $G$ with respect
to $\theta$ . If $G$ is CP-odd, we have
\begin{eqnarray}
  G_Q^{\mathrm{odd}}
  &=&G^{(1)}(0)\frac{iQ}{\chi_t V}\left(1-\frac{c_4}{2\chi_t^2 V}\right)
  +G^{(3)}(0)\frac{i Q}{2\chi_t^2 V^2}
  +O(V^{-3}) .
\label{eq:odd}
\end{eqnarray}

The formula (\ref{eq:even}) provides an estimate of the finite size effect
due to the fixed topological charge. The leading correction is of
order $O(1/V)$. 
It should be also noted that the other formula (\ref{eq:odd}) suggests
that it is possible to extract the $\theta$ dependence of CP-odd
observables, such as the neutron electric dipole moment by measuring 
the observable at a fixed non-zero topological charge, once the
topological susceptibility $\chi_t$ is obtained. 
%
%%%%%%%%%%%%%%%%%%%%%%%
\section{Topological susceptibility}
\label{sec:topological_susceptibility}
There are two reasons for measuring topological susceptibility.
The primary reason is to study whether the local topological
fluctuation is sufficiently created in unquenched QCD simulations. 
In conventional quenched QCD simulations, the topology change during
Monte Carlo updates is triggered by the formation of dislocations
which grows into local topological fluctuation with positive or
negative topological charge. This is a topology non-conserving
processes through lattice 
artifact.  On the other hand, in unquenched simulations such process
is highly suppressed towards continuum and chiral limit. In
particular, unquenched simulations by the JLQCD collaboration
explicitly prohibits such 
processes by introducing extra Wilson
fermions~\cite{Fukaya:2006vs}. Then, the main source 
of the local  topological fluctuation is achieved through the pair
creation  of local fluctuation of positive and negative topological
charge densities, just as in the continuum theory. Therefore, the
measurement of the topological susceptibility  is a crucial test of
the thermal equilibrium in local topological fluctuation. 
The secondary reason is that the topological susceptibility is the key
quantity to estimate the finite size effects at fixed topology as was
shown in the previous section.

Recently, there are several proposals for the field theoretical
definitions of the topological susceptibility, which are free from
ambiguities. The first one is given by Giusti et
al.~\cite{Giusti:2004qd} in which they 
define the topological susceptibility by the integration of
disconnected flavor-singlet pseudoscalar correlator.
%
%\begin{eqnarray}
%\chi_t \equiv \sum_x \langle  m P(x)  m P(0) \rangle_{\rm disc}.
%\end{eqnarray}
%
The second one is a UV divergence free definition by
Luscher~\cite{Luscher:2004fu} in terms
of n-point function of  flavor non-singlet scalar and pseudoscalar
correlator.
%, for example,  
%
%\begin{eqnarray}
%\chi_t \equiv m^5 \sum_{x_1,x_2,x_3,x_4} 
%\langle  \langle P_{12}(x_1) S_{23}(x_2) S_31(x_3)\rangle_F  
%         \langle P_{45}(x_4) S_{54}(0)\rangle_F  \rangle_A.
%\end{eqnarray}
%
The third one is proposed in the study of Schwinger
model~\cite{Fukaya:2004kp}. They extracted the topological
susceptibility from the asymptotical 
value of the singlet pseudoscalar correlator up to $1/V$ correction as 
\begin{eqnarray}
\lim_{x\rightarrow\infty} 
 \langle mP(x) mP(0)\rangle_{Q,V} 
= \frac{1}{V}\left( \frac{Q^2}{V}-\chi_t \right)
+ {\cal O}(V^{-2}).
\label{eq:PP}
\end{eqnarray}
Although the intuitive picture for this relation was given in
Ref.~\cite{Fukaya:2004kp}, the field theoretical proof based was given only
recently, which will be explained in the next subsection.
%
%%%%%%%%%%%%%%%%%%%%%%%%%%
\subsection{Field theoretical proof of the formula for the 
topological susceptibility}
Suppose that there is a well-defined local operator $\omega(x)$ 
that measures the local topological charge.
The global topological charge $Q$ is then obtained as
$Q =\int d^4 x\,\omega(x)$, and the topological susceptibility is
$\chi_t = \int d^4 x \langle\omega(x)\omega(0)\rangle$, where the expectation
value is taken for the $\theta=0$ vacuum.
Since $\omega(x)\omega(0)$ is CP-even, Eq.~(\ref{eq:even}) gives
\begin{eqnarray}
  \label{eq:bosonic_two-point}
  \langle \omega(x) \omega(0) \rangle_Q &=&
  \langle \omega(x) \omega(0) \rangle +
  \langle \omega(x) \omega(0) \rangle^{(2)} 
  \frac{1}{2V\chi_t} 
  \left(1 - \frac{Q^2}{V\chi_t}-\frac{c_4}{2\chi_t^2 V}\right)
\nonumber\\
& &
  +\langle \omega(x) \omega(0) \rangle^{(4)} \frac{1}{8\chi_t^2V^2}
  +O(V^{-3}),
\end{eqnarray}
where $\langle O \rangle^{(n)}$ is the $n$-th derivative 
of $\langle O \rangle$ with respect to $\theta$.
In the large separation limit $\vert x\vert\rightarrow\infty$, the CP
invariance at $\theta=0$ and the clustering property at a fixed $\theta$
gives 
\begin{equation}
  \label{eq:bosonic_two-pint_final}
  \lim_{\vert x\vert\to \mathrm{large}}
  \langle \omega(x) \omega(0) \rangle_Q =
  \frac{1}{V}\left(\frac{Q^2}{V}-\chi_t-\frac{c_4}{2\chi_t V} \right)
  +O(V^{-3}) + O(e^{- m_{\eta^\prime}\vert x\vert}),
\end{equation}
where the flavor singlet pseudo-scalar meson mass, $m_{\eta^\prime}$, is 
the lightest mass of possible intermediate states.  

Physical quantities such as the topological susceptibility $\chi_t$ can be
obtained through (\ref{eq:bosonic_two-pint_final}).
In practice, this formula will be used for a finite separation $x$ instead of
$|x|\to\infty$.
The clustering property in the $\theta$ vacuum receives
a correction of order of $e^{-m_{\eta'}|x|}$, which vanishes quickly because
the flavor singlet meson $\eta'$ acquires a large mass due to the axial
anomaly of QCD.

We now express the bosonic correlation function
$\langle\omega(x)\omega(0)\rangle$ in terms of a fermionic one using
the anomalous axial U(1) Ward-Takahashi (WT) identities for 
an arbitrary operator $O$:  
\begin{equation}
  \label{eq:WT}
  \langle \partial_\mu A_\mu (x) O -  2m P(x) O +2\omega(x) O +
  \delta_x O \rangle = 0 ,
\end{equation}
where 
$A_\mu(x) =\frac{1}{N_f}\sum_f\bar\psi^f(x)\gamma_\mu\gamma_5\psi^f(x)$ 
and $P(x) = \frac{1}{N_f}\sum_f\bar\psi^f(x)\gamma_5\psi^f(x)$ are
the flavor singlet axial-vector current and pseudo-scalar density,
respectively,
and $\delta_x O$ denotes a axial rotation of the operator $O$ at $x$.
Combining WT identities for $O=2mP(0)$ and $O=2\omega(x)$ 
Combining these we finally obtain
\begin{equation}
  \lim_{\vert x\vert\to \mathrm{large}}
  \langle m P(x) m P(0) \rangle_Q
  = 
  \frac{1}{V} \left(\frac{Q^2}{V}-\chi_t - \frac{c_4}{2 \chi_t V}\right) 
+ O(e^{-m_{\eta^\prime}\vert x\vert}).
  \label{eq:singlet}
\end{equation}
%
%\begin{figure}[here]
%\includegraphics[width=.5\textwidth]{chit_mq_Q0.eps}
%\includegraphics[width=.4\textwidth]{etap_m0025.eps}
%\caption{Numerical study of the topological susceptibility.
%Figures from T.W.Chiu's talk.}
%\label{fig:Chiu}
%\end{figure}
%
In fact, using this formula one can 
determine the topological susceptibility 
from the nonzero asymptotic value of the singlet pseudoscalar
correlator~\cite{Aoki:2007pw, Chiu}.
% as shown in Fig.~\ref{fig:Chiu}.
%
%
%%%%%%%%%%%%%%%%%%%
\section{Application to physics}

\begin{figure}[tb]
\hspace{-0.5cm}
\includegraphics[width=.6\textwidth]{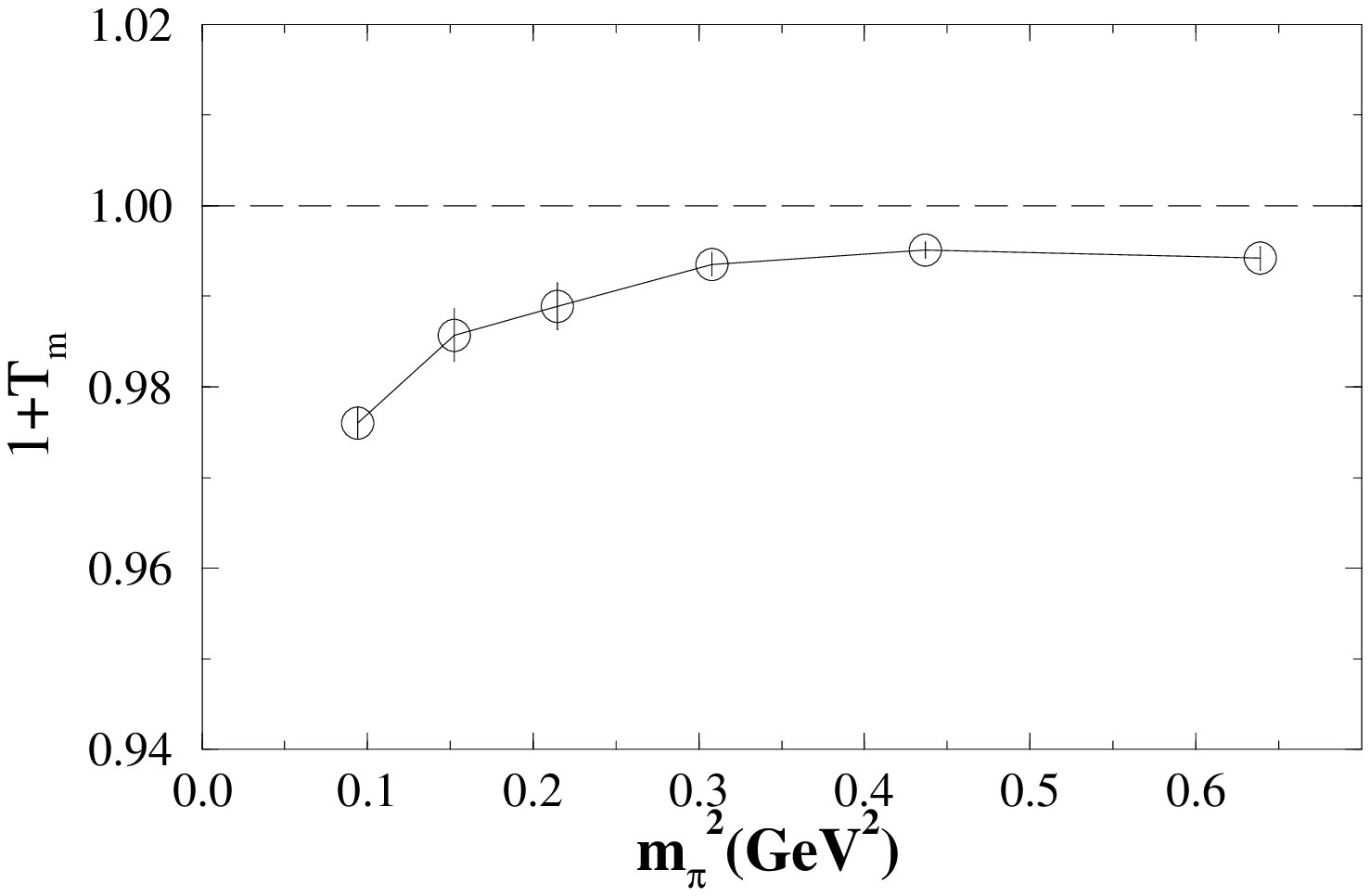}
\hspace{-0.5cm}
\includegraphics[width=.6\textwidth]{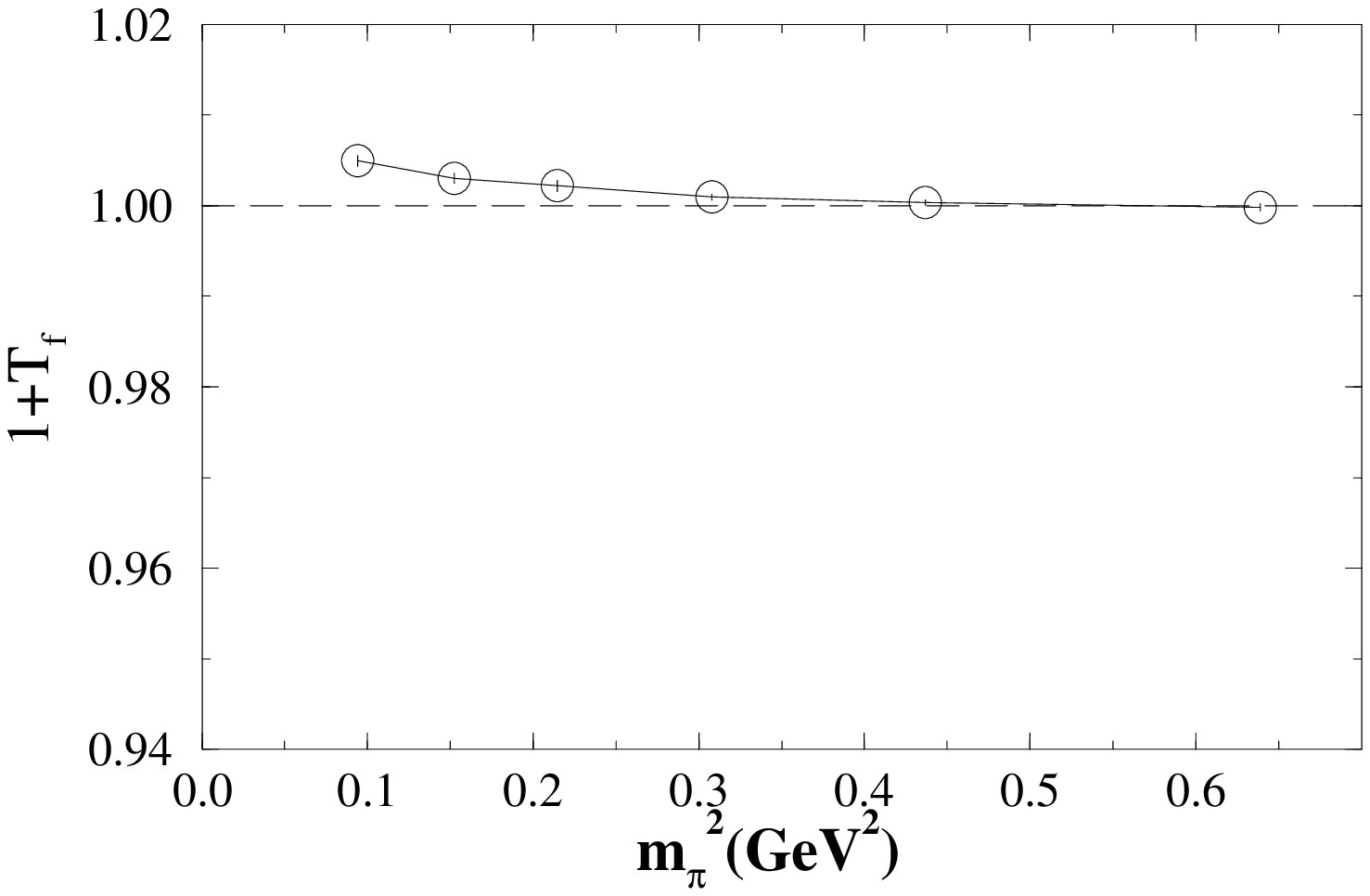}
\caption{
The finite size effects $1+T_m$ (left panel) and $1+T_f$ (right panel)
from fixed topology at $Q=0$ with $L=2$ fm. 
The pion masses correspond to those in the  dynamical
simulation by the JLQCD collaboration. }
\label{fig:FSE}
\end{figure}
%
%\vspace{-0.2cm}
One can estimate the finite size corrections for the pionic quantities
with the help of Chiral Perturbation Theory (ChPT). Using the
next-to-leading order ChPT formula, $\theta$ dependence of the the
pion mass and the decay constant for two-flavor QCD is
\begin{eqnarray}
m_{\pi}^2(\theta)|_{rm NLO} 
&=& m_{\pi}^2(\theta)
 \left[ 
  1 + \left(\frac{m_{\pi}(\theta)}{4\pi f}\right)^2
 \left( 
  \ln \left(\frac{m_{\pi}(\theta)}{m_{\pi}^{\rm phys}}\right)^2 
 - \bar{l}_3^{\rm phys} \right)\right],
\\
f_{\pi}(\theta)|_{rm NLO} 
&=& f  
\left[ 1 - 2 \left(\frac{m_{\pi}(\theta)}{4\pi f}\right)^2
 \left(\ln \left(\frac{m_{\pi}(\theta)}{m_{\pi}^{\rm phys}}\right)^2 
 - \bar{l}_4^{\rm phys} \right)\right],
\end{eqnarray}
where 
$m_{\pi}^2(\theta)\equiv 2 B_0 m_q
\cos\left(\frac{\theta}{N_f}\right)$, 
and $\bar{l}_3$, $\bar{l}_4$ are the low energy constants which can be
estimated as $\bar{l}_3^{\rm phys} =2.9 \pm 2.4$, and $\bar{l}_4^{\rm
phys} =4.4 \pm 0.2$~\cite{Colangelo:2001df}.
Fig.~\ref{fig:FSE} shows the finite size effects $1+T_m\equiv
m^{Q=0}_{\pi}/m_{\pi}$ and $1+T_f\equiv
f^{Q=0}_{\pi}/f_{\pi}$ at $L=2$ fm using
the NLO ChPT. The pion masses correspond to those in the  dynamical
simulation by the JLQCD collaboration~\cite{Noaki:2007es}. The
topological susceptibility 
$\chi_t$ is extracted from the singlet pseudoscalar correlator as
explained in the previous section. It is found that finite size effect
for the pion mass ranges from 0.5 to 2.5 \%, whereas that for the
pion decay constant is well below 0.5\%. These finite size correction
from  fixing the topology is correctly taken into account in the
spectrum study~\cite{Noaki:2007es}. In general, for quantities which
has a non-vanishing 
chiral limit, $\theta$ or Q dependence correction only comes through
$m_{\pi}(\theta)$ as sub-leading corrections. Therefore, pion receives
the largest finite size correction. This means that if the finite size
effect of the pion mass in under control, other hadronic quantities are safe.
%
%%%%%%%%%%%%%%%%%%%%%%%
\subsection{Nongaussianity of the topological charge distribution}
Recently deviation of the topological charge distribution from
Gaussian is observed for pure Yang-Mills gauge theory
~\cite{Del Debbio:2002xa,D'Elia:2003gr, Giusti:2007tu,Del Debbio:2007kz}.
In principle, we can also measure the deviation from the Gaussian
($c_4$ coefficient) by combining the 2-, 3-, 4-point functions of the 
topological charge density given as follows
\begin{eqnarray}
  \lim_{|x|\to\mathrm{large}}
  \langle\omega(x)\omega(0)\rangle_Q
  &=&
  -\frac{\chi_t}{V^2}
  \left[1- \frac{1}{2\chi_t^2 V}(c_4-2\chi_t Q^2)\right] + 
  O(V^{-3})
\\
  \lim_{|x_i-x_j|\to\mathrm{large}}
  \langle\omega(x_1)\omega(x_2)\omega(x_3)\rangle_Q
  &=&
  -3\chi_t\frac{Q}{V^2}
  \left[1+\frac{7}{6\chi_t^2 V}(c_4-\frac{2}{7}\chi_t Q^2)\right] + 
  O(V^{-4})
\\
  \lim_{|x_i-x_j|\to\mathrm{large}}
  \langle \omega(x_1) \omega(x_2) \omega (x_3) \omega(x_4)\rangle_Q
  &=& 3\frac{\chi_t^2}{V^2}\left[1+\frac{1}{\chi_t^2 V}(c_4-\chi_t
  Q^2)\right]^2  
  +O(V^{-4}).
\end{eqnarray}
\section{Summary}
\label{sec:summary}
We have derived general formulas which express arbitrary
correlation functions at a fixed topological charge $Q$ in terms of
the same correlation function (and its derivatives) in the $\theta$ vacuum.
The difference between the fixed $Q$ vacuum and the fixed $\theta$ vacuum
can be shown to disappear in the large volume limit as $1/V$ only
using fundamental properties of the quantum field theory.

These formulas open a new possibility to calculate physical quantities in the
lattice QCD simulations at a fixed topological charge.
This will become unavoidable as the continuum limit is approached,
irrespective of the lattice fermion formulation one employs,
as far as the algorithm is based on the continuous evolution of the gauge
field. 

Applying the formula for $n$-point functions of the topological
charge density $\omega$, we have shown that 
the topological susceptibility $\chi_t$ and $c_4$ appear 
at the first and second order corrections, respectively.
In principle, these parameters can be determined by the lattice data.
Our method is free from short-distance singularities, since the local
topological charge operators are put apart from others and no contact 
term appears. 
Numerical calculation is in progress by the JLQCD collaboration on the gauge
configurations generated with dynamical overlap fermion
~\cite{Kaneko:2006pa,Hashimoto:2006rb,Matsufuru:2006xr,Yamada:2006fr}.
Once these parameters are numerically obtained, they can be used as input
parameters for other physical observables, such as weak matrix
elements such as $B_K$~\cite{Yamada:2007nh}, pion form factor
~\cite{Kaneko:2007nf}, the neutron electric dipole moment, and so on.  
\section*{Acknowledgments}
We would like to acknowledge the workshop at Yukawa Institute  
YITP-W-05-25 ``Actions and Symmetries in Lattice Gauge Theories,''
where part of this work was initiated.
This work is supported in part by the Grants-in-Aid for
Scientific Research from the Ministry of Education,
Culture, Sports, Science and Technology.
(Nos. 13135204, 15204015, 15540251, 16028201, 18034011, 18340075, 18840045,
19540286). 
%
%
%=====================

%
%
%
%
\end{document}